\documentclass[10pt]{article}

\usepackage{amsmath}
\usepackage{amssymb}

\usepackage{cite}

\usepackage{lineno}

\usepackage{graphicx}
\usepackage{float}
\usepackage{caption}
\usepackage{subcaption}
\DeclareGraphicsExtensions{.pdf,.png,.jpg,.eps,.tiff}

\usepackage{booktabs, multirow, tabularx}
\usepackage[table]{xcolor}
\definecolor{lightgray}{gray}{0.9}
\usepackage{microtype}
\DisableLigatures[f]{encoding = *, family = * }

\usepackage[labelfont=bf,labelsep=period,justification=raggedright]{caption}

\makeatletter
\renewcommand{\@biblabel}[1]{\quad#1.}
\makeatother

\date{}

\begin{document}

\begin{flushleft}
{\Large
  \textbf{Vaccination and public trust: a model for the dissemination of vaccination behavior with external intervention.}
}
\\
Andr\'es D. Medus$^{1,\ast}$, 
Claudio O. Dorso$^{1}$
\\
\bf{1} {Departamento de F\'isica and IFIBA - CONICET, Facultad de Ciencias Exactas y Naturales,
 Universidad de Buenos Aires, Pabell\'on 1, Ciudad Universitaria, Ciudad Aut\'onoma de Buenos Aires (1428), Argentina}
\\
$\ast$ E-mail: admedus@df.uba.ar (ADM)
\end{flushleft}

\section*{Abstract}
Vaccination is widely recognized as the most effective way 
of immunization against many infectious diseases.
However, unfounded claims about supposed side effects of some 
vaccines have contributed to spread concern and fear among people, thus inducing vaccination refusal.
For instance, MMR (Measles, Mumps and Rubella) vaccine coverage has undergone an important decrease in a large part of Europe and US
as a consequence of erroneously alleged side effects, leading to recent measles outbreaks.  
In this work, we propose a general agent-based model to study the spread of vaccination behavior
in social networks, not as an isolated binary opinion spreading on them, but as part of a process 
of cultural dissemination in the spirit of Axelrod's model. We particularly focused on the impact of a small anti-vaccination movement over
an initial population of pro-vaccination social agents.  
Additionally, we consider two classes of edges in the underlying social network: 
personal edges able to spread both opinions and diseases; and the non-personal ones representing interactions mediated by information technologies, 
which only allow opinion exchanges.
We show that clusters of unvaccinated agents emerge as a dynamical outcome of the model, increasing 
the probability of occurrence and the final size of measles outbreaks. 
We illustrate the mitigating effect of a public health campaign, represented by an external field, 
against the harmful action of anti-vaccination movements.
Finally, we show that the topological characteristics of the clusters of unvaccinated agents determine the scopes of this mitigating effect.


\section*{Introduction}

Vaccination constitutes one of the main ways to prevent the spread of infectious diseases in modern era only behind safe water \cite{Plotkin2004}. 
It is largely recognized as the most effective
method for immunization, with great success in the worldwide eradication of smallpox and in the control of other infectious diseases such as 
measles, rubella, tetanus, and polio almost all over the world \cite{WHO2008,Maurice2009}. 
However, recent outbreaks of measles in UK and US have awoken renewed concern about vaccination rejection \cite{Wales2013,US2014}.
In the particular case of MMR vaccine against measles, mumps and rubella, quickly refuted claims about a supposed link with autism 
\cite{Wakefield1998,Taylor1999} have aroused distrust and fear in people. Thus, people's perception of vaccine safety and efficiency
  has shown to be an important factor for vaccine uptake that, at the same time, can spread between individuals \cite{Burgess2006,Brown2012}.

In those places where vaccination is not mandatory, to vaccinate or not vaccinate a child is a parent's decision 
which usually involves cultural and belief systems. 
However, decisions are not made by individuals as isolated entities, 
but as social agents interacting with other social actors.
In this regard, recent works have begun to model the dissemination of vaccination opinion between peers 
as a pure social imitation process in a mean field approach \cite{Bauch2005}. 
As an extension of these previous models, some authors have also introduced vaccine side effects
to be assessed by social agents, together with contagion risks, in a game-theoretical framework \cite{dOnofrio2011}.   

Mean field models usually assume that agents' opinions evolve in a homogeneous mixing approach, by which each agent 
can interact with every other agent.
A step forward in the description of opinion dissemination with social contact heterogeneities is provided 
by agent-based modeling (ABM) where opinions spread over a social network.
The clustering of unvaccinated, and then susceptible, agents constitutes one of the most important outcomes from this novel approach 
\cite{Salathe2008,Meyers2012,Salathe2014,Liu2013}.
The clustering of susceptible agents leads to an effective shift of the safety threshold obtained by homogeneous herd immunity 
(i.e. the expected herd immunity under the assumption of random mixing), thus increasing the chances of disease outbreaks.

Today we know that opinions not only spread by personal interaction, but also through new ways of remote-socialization provided
by current information technologies. 
Opinions and even sentiments can be spread by virtual friendship networks, as some authors have shown in recent works \cite{Bond2012,Fowler2014}.
However, current ABM models to study the dissemination of vaccination behavior assume 
that opinions spread over exactly the same personal social ties subsequently involved in disease spreading.

On the other hand, ABM models for dissemination of vaccination behavior in social networks 
consider homogeneity at the level of social interactions, by which all social ties are supposed to be equally influentials. 
However, opinions mostly spread between similar individuals, as has been recently shown for the particular case of health behavior \cite{Centola2011}.
For instance, a given individual could adopt the vaccination behavior of one of his neighbors if both have a previous degree of affinity. 
The preferential interactions between similar individuals is a well-known concept in sociology, called \textit{homophily} 
\cite{Lazarsfeld1954,McPherson2001}. 
In this context, homophily acts as a weight for the ties between social agents. 
This last assumption constitutes the cornerstone of Axelrod's model, proposed in order to explain how social consensus arises 
in human societies \cite{Axelrod1997}.
In the particular case of vaccination opinion, we are interested in the opposite transition: 
how does the initially broad consensus about, for instance, MMR vaccine in developed countries with a wide vaccine availability, such as UK and US \cite{WHO2014cov}, 
begin to decline due to the action of small anti-vaccination movements?

Recent outbreaks of vaccine-preventable diseases have shown that although vaccine availability constitutes a necessary condition to 
reach global immunization, it is not sufficient. 
In this regard, massive public health campaigns can contribute to mitigate vaccination refusal by spreading information about
the benefits of vaccination. However, as far as we know, this subject has not received much attention in 
previous models accounting for social network structure, beyond a few exceptions based on mean field approach \cite{Manfredi2012}.

In this work, we explore the effects of opinion spreading by a small group of anti-vaccination stubborn individuals, initially immersed in a
totally pro-vaccination population with social network structure. 
We avoid contagion risk perception in social agents by considering a disease with very low initial incidence, due to the vaccine's effectiveness, 
compatible with the case of measles for the early years after MMR vaccine introduction in developed countries. 
In this way, we argue that the vaccination behavior of social agents only depends
on the influence exerted by peers and other social actors through a process of opinion dissemination.

Two classes of edges are considered as part of the underlying social network: a) opinion-exchange edges without direct personal interaction,
and b) direct personal edges which are also involved in the opinion dissemination process but, in turn, are able to transmit an infectious disease. 
These different kinds of edges define two entangled subnetworks over which the process of opinion dissemination takes place.

We propose an extension of Axelrod's model in order to study the spread of vaccination behavior as 
one of the features involved in the cultural exchange between social agents. 
Cultural exchange occurs in two different ways. On one side, as a static adaptation process where social agents exchange their opinions
with their fixed neighbors in order to reach consensus. On the other side, as an active adaptation process where each social agent, in addition 
to the static opinion adaptation, can also replace his dissimilar neighbors by others with which he has higher homophily degree \cite{Centola2007}.
We aim to focus on the dynamical basis of clusterization of unvaccinated agents, by placing particular emphasis on the 
topological features of these clusters and their relation with the spread of an infectious disease. We consider measles as case study. 

We compare outbreak sizes of measles over the unvaccinated population yielded by opinion dissemination models, with those obtained
assuming that the same unvaccinated population is distributed at random over the social network. This comparison is also proposed for the particular
case satisfying the recommended $95\%$ MMR vaccine coverage \cite{Plan2012}. 
Finally, we study the effects of massive public campaigns undertaken by health authorities in order to mitigate the action of anti-vaccination activists.

\section*{Methods}

We will discuss the main elements of our model to formulate its final algorithmic description at the end of the present section.

\subsection*{Social structure}\label{structure}
We model a structured population through a network representation, where nodes and edges are respectively associated
to social agents and their social ties. In this context, a social agent may represent a family or an individual agent. Here 
we refer to a social agent as a family composed by immune adults who must decide about vaccination, or not, of their only son or daughter. Thus, the
social ties account for all the social interactions among the members of different households.
It is important to distinguish those personal interactions by which 
an infectious disease could be transmitted, from those which, although having impact on individuals' opinions, 
do not imply a direct personal interaction. 
With this in mind, we consider two classes of social edges: a) direct personal and b) opinion-exchange non-personal edges. Personal edges come from 
strong, frequent or close contacts implying spatial proximity. They allow not only opinions exchange but also the spread of an infectious disease. 
Some examples are family ties, or those to close friends, coworkers or neighbors. 
On the other side, non-personal opinion-exchange edges may represent weak, distant or indirect relations, 
such as friendship relations in online social networks or ``following'' relationships in micro-blogging platforms.
They represent those social contacts that, in spite of not having spatial proximity, are able to disseminate opinions. 
As a consequence, the social structure can be represented as two entangled sub-networks, one containing all the direct personal ties, 
while the other one considering exclusively non-personal opinion-exchange ties. 

In order to simplify the representation, we initially assume a square network plus second neighbors ties (3N+2N square network), 
with $N=2500$ nodes ($50\times 50$ square network) and with homogeneous degree $k=16$, as represented in Figure~\ref{fig1}. 
In addition, as shown in Figure~\ref{fig1}, each node has $k_p=8$ neighbors corresponding to direct personal contacts and $k_o=8$ non-personal contacts.

\subsection*{Opinion dissemination model}\label{opinion}
We adopt a novel variant of the well-known Axelrod's model for cultural dissemination in social networks. 
In particular, we aim to model the spread of vaccination behavior in the previously described social network. 
However, this is only one of the features involved in a plentiful cultural exchange between social agents. All the other features evolve in an 
adaptation process by which the degree of homophily between connected social agents is dynamically developed. 
At the same time, the degree of homophily acts as a ``catalyst'' for the spread of vaccination behavior between social agents. 

In their original formulation, Axelrod's model considers a population of interacting social agents with cultural background described 
through a number of features $F\in \mathbb{N}$, representing a particular belief system, involving different subjects such as politics, education, sports, entertainments, etc.
Each of such features admits a multiplicity of traits $q \in \mathbb{N}$.
We adopt the Axelrod's model with $F+1$ features, and a heterogeneous number of traits $q$.
Vaccination-related opinion is a binary trait ($q=2$), i.e., only two opposing traits are admitted: vaccinator or non-vaccinator.
The remaining $F$ features are assumed to have uniform $q$, with $q\geq 2$. Thus, the cultural background of a given social agent $i$
can be mathematically described by a time-dependent state vector $\mathbf{V}^i(t)=(V^i_1(t),V^i_2(t),...,V^i_{F+1}(t))$ 
with $\mathbf{V}^i(t)\in \mathbb{I}^{F}\times \{0,1\}$, where $\mathbb{I}=\{1,2,...,q\}$ and $V^i_{F+1}(t)=0$ corresponds to pro-vaccination opinion while 
$V^i_{F+1}(t)=1$ to anti-vaccination opinion at time $t$. 
Then, we say that two social agents $i$ and $j$ agree in a particular
feature $n$ at time $t$ when $V^i_{n}(t)=V^j_{n}(t)$. In addition, we define the homophily degree $h(i,j)\in \mathbb{N}_0$ between an agent $i$ and one 
of his neighbors $j$, as the total number of features in which they agree.
At this point, we must remember that, in the context of our model, each social agent does not represent an individual but a 
family group. Thus, we will assume only one opinion vector $\mathbf{V}^i(t)$ representing the decision maker's opinions within each family group.

Essentially, the opinion dissemination model consists of social agents performing an adaptation process with asynchronous update. As initial condition, 
 $\mathbf{V}^i(t=0)$ are assigned uniformly at random in $\mathbb{I}^{F}\times \{0,1\}$ for all $i\in \{1,...,N\}$.
In each time step, a randomly chosen agent $i$ will interact with one of his neighbors $j$, also chosen uniformly at random, with a probability $P(i\rightarrow j)$ 
proportional to $h(i,j)$. Both personal and non-personal neighbors are involved in the opinion dissemination process.
In addition, we define an homophily threshold $\kappa$ above which the interaction 
takes place. Then, $P(i\rightarrow j)$ is written as:
\begin{equation}\label{eq1}
 P(i\rightarrow j)=\left\{\begin{array}{lll}
              0 & \text{if} & h(i,j)<\kappa\\
             \frac{h(i,j)}{F+1} & \text{if} & h(i,j)\geq \kappa \hspace{0.5cm}.
            \end{array}\right.
\end{equation}
When the social interaction becomes effective, $i$ adopts the opinion of $j$ for a randomly chosen feature in which they previously disagree, i.e., 
 if ${V}^i_n(t-\delta t)\neq{V}^j_n(t-\delta t)$, then ${V}^i_n(t)={V}^j_n(t)$. It is worth emphasizing that vaccination opinion stands at equal footing 
 with all the other features. 
In the context of the model here proposed, we choose $F=10$ and $\kappa=2$ to avoid social interactions where the opinion about 
vaccination was the only coincidence. This last assumption implies that vaccination opinion would be spreading in virtue of previous cultural affinity between social agents.

In order to consider the action of anti-vaccination movements, we select a fraction of stubborn agents $p_{stb}$ from the total population.
Stubborn agents ($S_{stub}$) preserve their vaccination opinion fixed ${V}^i_{F+1}(t)=1$ for all time $t$, while their opinions about the other
features are subject to the adaptation process. 
Thus, they can influence their neighbors about vaccination behavior, but can not be influenced by their neighbors in this respect.
At time $t=0$, they are placed homogeneously at random in the network. Meanwhile, all the other
non-stubborn agents are initially vaccinators (${V}^i_{F+1}(0)=0$), because we are interested in studying the impact of anti-vaccination movements 
in the spread of vaccine rejection behavior.

\subsection*{Adaptive network}\label{rewiring}
 In social sciences, there are essentially two main mechanisms to explain why related individuals share traits (opinions): 
 i) \textit{social imitation}, and ii) \textit{linking by homophily} \cite{Aral2009,Fowler2013}. Social imitation is the mechanism of adaptation 
 by which social agents adopt, or imitate, the traits of their neighbors. For this reason, social imitation constitutes a \textit{static adaptation} 
 process because it does not involve structural changes in the underlying social network.
 On the other hand, the alternative mechanism, which we called \textit{linking by homophily}, assumes that each social agent
 performs preferential connections, in an \textit{active adaptation} process guided by his degree of homophily with the other social agents.

We propose two alternative models regarding the dynamic of edges: 
\begin{itemize}
 \item[a. ]the \textit{static network} model, described in the previous section\ref{structure}, 
 for which every social agent only perform a passive adaptation process mediated by homophily, for which both personal and non-personal edges result static and permanent;
 \item[b. ]the \textit{adaptive network} model, for which social agents follow an active
 adaptation process, in addition to the previously described static adaptation, involving an edge rewiring mechanism. 
 In this way, both the social structure and the process of opinion adaptation evolve together.
\end{itemize}

The rewiring process of \textbf{b.} comprises two alternative paths depending on whether the chosen edge is personal or non-personal. 
Both cases are outlined below, in what follows $i$ represents the source agent and $j$ the target agent:
\begin{itemize}
 \item[i. ] Non-personal edge $l(i,j)$: $i$ will attempt to rewire the edge to another agent $k$ chosen uniformly at random
(supposing that $k$ was not previously linked to $i$). If $P(i\rightarrow j)<P(i\rightarrow k)$, the rewiring will be accepted, otherwise, 
it will be refused and substituted by a passive opinion adaptation attempt between $i$ and $j$.
 \item[ii. ] Re-wireable personal edge $l(i,j)$: $i$ will attempt a rewiring step to another agent $k$ chosen between their non-personal
 neighbors. If $P(i\rightarrow j)<P(i\rightarrow k)$, the edge $l(i,k)$ becomes a personal one, whereas $l(i,j)$ also changes its
 status becoming a non-personal edge.
\end{itemize}
These last prescriptions preserve the total number of personal and non-personal edges and, as a direct consequence, do not alter
the average degrees $\langle k_o \rangle$ and $\langle k_p \rangle$. Furthermore, the particular prescription for rewiring of personal edges 
is aimed to reproduce what we expect in real situations. That means that non-personal relations help to develop trust between social 
agents and, later on, might become personals.

It is in the adaptive feature of social networks where the distinction between personal and non-personal edges becomes evident.
By definition, personal edges are robust and, then, hardly prone to active adaptation.
In contrast, non-personal contacts are essentially volatile.
Thus, we assume that all non-personal contacts can be rewired, but this can only occur for a small fraction $p_{pc}$ of personal edges chosen
at $t=0$ and remaining thereafter as re-wireable edges.
The subgraph $G_{pers}$, defined only by personal contacts, should be more robust than the corresponding non-personal subgraph, but
it is also desirable that satisfies \textit{small-world} conditions \cite{Watts1998} in order to reproduce one of the crucial topological features of real social networks.
For these reasons, we choose a small value of $p_{pc}=0.1$ for the case of the adaptive network model (for other values $0<p_{pc}<0.1$ we have observed the same qualitative results).

\subsection*{External field}
The public health intervention in the dissemination of vaccination behavior can be represented in the form of an external field $\phi$.
All the social agents are exposed to $\phi$, but in contrast with the vector field proposed in previous works \cite{Avella2005,Avella2006}, 
here it only acts as a bias for vaccinator-to-non-vaccinator transition. 
In other words, $\phi$ represents a public health information campaign in order to prevent vaccinator-to-non-vaccinator transitions. Thus, $\phi$ is involved in the 
opinion dissemination process when a vaccinator source agent $i$ has an effective opinion interaction with a non-vaccinator target $j$,
and vaccination behavior is chosen for interaction. In this case, if $V^i_{F+1}(t)=0$ and $V^j_{F+1}(t)=1$, the resulting social imitation process
is:
\begin{equation}\label{eq2}
 V^i_{F+1}(t+\delta t)=\left\{\begin{array}{lll}
               V^j_{F+1}(t) & \text{with probability} & (1-\phi)\\
              V^i_{F+1}(t) & \text{with probability} & \phi \hspace{0.5cm}. 
            \end{array}\right .
\end{equation}
As it is clear from Equation \ref{eq2}, the classical social imitation process is recovered by replacing $\phi=0$.

\subsection*{The model}
Here we gather all previous processes in the final algorithmic structure. We begin with all social agents placed on the regular social network
of Figure~\ref{fig1}. Initially, the vector states $\mathbf{V}^i(t=0)$ are assigned uniformly at random for all $i\in \{1,...,N\}$, with the exception
of vaccination behavior which is initially set at $V^i_{F+1}(t=0)=0$, unless the agent $i$ is a stubborn agent ($i\in S_{stub}$) for which $V^i_{F+1}(t=0)=1$
at all time $t$.
The algorithm proceeds as follows:
\begin{enumerate}
 \item A random source agent $i$ is selected uniformly at random, and a target agent $j$ is selected between the neighbors of $i$.
 \item Static network model: $i$ follows the opinion adaptation process described in {Opinion dissemination model}~\ref{opinion}. If $i\in S_{stub}$ (is a stubborn agent) then it can
 not adapt their vaccination opinion. Go to step 4.
 \item {Adaptive network model: 
 \begin{itemize}
  \item[i. ] $i$ and $j$ are joined by a re-wireable edge: $i$ follows the prescription of the edge rewiring process described in {Adaptive network section}~\ref{rewiring}.
  \item[ii. ] $i$ and $j$ are joined by a personal non-rewireable edge: $i$ follows the opinion adaptation process described in {Opinion dissemination model}~\ref{opinion}. 
  If $i\in S_{stub}$ (is a stubborn agent) then it can not adapt their vaccination opinion.
 \end{itemize}}
 \item {Action of the external field: if $\phi>0$ and the vaccination opinion of $i$ has changed in previous steps from $V^i_{F+1}(t-1)=0$ to $V^i_{F+1}(t)=1$ by the 
 opinion adaptation process, then $i$ becomes vaccinator again ($V^i_{F+1}(t)=0$) with probability $\phi$. In other words, the vaccinator-to-non-vaccinator transition is aborted with
 probability $\phi$.}
 \item{Stop condition: the algorithm stops when the system reaches a metastable configuration (transitions are not longer possible), 
 or a partially frozen configuration without considering small intermittent fluctuations 
 in the vaccination opinion due to the action of stubborn agents (frozen configuration for all features, except vaccination opinion).}
 \item {$t=t+\delta t$. Return to step 1.}

\end{enumerate}

\section*{Results and Discussion}

Setting $q=170$ (corresponding to the critical value $q_c$ for the model without external field on the static network, as shown in Figure Supp1~\ref{figS1}), 
we have obtained alternative behaviors for the size distribution of the unvaccinated population at the end of the opinion dissemination process 
in both the static and the adaptive models.
From a big picture perspective, we show in Figure~\ref{fig2} the strong dependence of the total unvaccinated population with $q$ and 
with the strength of the external field $\phi$, both for the static model (Figure~\ref{fig2}-A) as well as for the adaptive one (Figure~\ref{fig2}-B). 
It can also be observed in Figure~\ref{fig2} that the range of $q$-values differ in static and adaptive models due to the known Axelrod's 
model sensitivity to network topology \cite{SanMiguel2003}. Despite slight differences, the qualitative conclusion is the 
same in both cases: a moderate public campaign promoting vaccination has a strong impact at the level of individuals' opinion, 
consequently resulting in a large vaccine coverage. On the other hand, in absence 
of external field $\phi$, the cultural dissemination process with the intervention of an anti-vaccination movement
-representing only $1\%$ of the total population- can lead to a large increase in vaccine refusal.

Figure~\ref{fig3} shows some distributions of unvaccinated population ranging from extended U-shaped for $\phi=0$, to unimodal short-tailed for the case corresponding 
to an applied external field $\phi=0.02$. Again, the strong mitigating effect induced by $\phi$ is
emphasized by the sequence of histograms of unvaccinated population (Figure~\ref{fig3}). 
In order to quantify this assertion for an epidemic scenario (see Epidemiological model description~\ref{Supp} for details about the measles spreading simulations),
we define the right tail distribution $P(x\geq c)= 1 - P(x < c)$ representing the probability to obtain an outbreak of size equal or larger than $c$ as:
\begin{equation}\label{eq3}
 P(x\geq c)=\sum_{n\geq c}p(n)
\end{equation}
with $p(n)$ the probability to get an outbreak of size $n$. The final outcome of the external field action is represented in Figure~\ref{fig4}, where $P(x\geq 25)$
is shown for $\phi\in [0.00,0.01,0.02]$. Essentially, Figure~\ref{fig4} states that $\phi$ has greater impact on the adaptive network model. 
In absence of external field, $P(x\geq 25)$ is larger for the adaptive model than for the static one. In contrast, the opposite situation prevails 
 when the external field starts to grow up.
We will see later that this puzzling effect can be explained from the topological analysis of the clusters of unvaccinated agents.

\subsection*{Clustering of unvaccinated agents}
Figure~\ref{fig5} compares the mean outbreak size of measles as a function of the unvaccinated population obtained at the end of the opinion adaptation process,
with those which would be obtained if the same unvaccinated population were randomly distributed over the final social network, in what we called \textit{the percolation approach}.
An immediate conclusion is that both models give rise to larger measles outbreaks than those obtained considering the 
percolation approach, at least below a threshold size of unvaccinated population.

Given a population of unvaccinated agents, a particularly relevant aspect is to know how they are distributed 
over the network. Figure~\ref{fig6} shows the cluster size distribution corresponding to the unvaccinated distributions of Figure~\ref{fig3}.
In this context, a cluster is defined by considering the subgraph $G_{pers}$ involving only direct personal edges, i.e., those through which an infectious disease could be transmitted.
The cluster size distributions given by both models are compared with those obtained from the equivalent percolation processes over the same final networks
(for the case of the adaptive network model, the percolation process could be performed alternatively over the initial social network, deprived of 
the structural changes imposed by the active adaptation, what could be done by comparing the left and right panels of Figure~\ref{fig6}).  
As can be seen in Figure~\ref{fig6}, both models show significant differences with respect to their associated percolation results, in particular for large clusters which 
are much more frequent in both the static and the adaptive models. In order to quantify these comparisons we also show in the insets of Figure~\ref{fig5}
the values of the t-statistic $t_S(i)$ for each bin $b$, defined as:
\begin{equation}\label{eq4}
 t_S(b) = \frac{\rho(b)-\hat{\rho}_{perco}(b)}{\hat{\sigma}_{perco}(b)}
\end{equation}
being $\rho(b)$ the density corresponding to the bin $b$ given by the model, while $\hat{\rho}_{perco}(b)$ and $\hat{\sigma}_{perco}(b)$
are, respectively, the mean and the standard deviation of the sample given by the percolation realizations, also corresponding to the bin $b$.
Figure~\ref{fig6} reveals that the bigger values of $t_S(b)$ are mainly obtained for large cluster sizes. This result unveils
the tendency toward clusterization of unvaccinated agents in the context of the proposed models for the spread of vaccination behavior. 

As a partial conclusion, we show that the grouping or clusterization of unvaccinated individuals has direct consequences on the incidence of an infectious disease: the more
clusterized the unvaccinated are, the more infected agents we have at the end of an outbreak. Thus, herd immunity is strongly affected by the clustering
of susceptible agents.

\subsection*{High vaccine coverage}
Vaccine coverage usually reaches higher levels in developed countries with more vaccine availability. 
In particular, recommendations of WHO suggest $95\%$ vaccine coverage for the case of MMR vaccine
in order to eradicate measles. Thus, we have repeated our previous analysis but now filtering those realization leading to a final immunized population between $94.6\%$ to $95.4\%$ 
- equivalent to an unvaccinated population between $115$ to $135$ individuals over a total population of $2500$ - aiming to be close to WHO recommendation.
Figure~\ref{fig7} clearly shows that the probability to obtain large outbreaks are higher for both the static and the adaptive network models, when 
they are compared with the percolation approach. Again, this assertion can be quantified through the right tail distribution $P(x\geq \eta)$ defined in equation \ref{eq3}.
Taking $\eta=25$ ($1\%$ of total population $N$), we obtain $P^{\phi=0}(x\geq 25)=0.302$ 
for static model and $P^{\phi=0}(x\geq 25)=0.034$ for adaptive model, while $P^{\phi=0}(x\geq 25)=0$ for both equivalent percolation approaches. 
On the other hand, Figure~\ref{fig8} shows that larger clusters are again more frequent for both opinion dissemination models than for the corresponding percolation
approach, which confirms the previous statement linking the clustering of unvaccinated with higher outbreak probability and larger mean outbreak size. 
In this particular case, the effects of the clustering of unvaccinated are magnified when compared with the percolation approach. 
This magnifying effect is rooted in the sub-critical percolation regime imposed by the small unvaccinated population.

\subsection*{Clusters topology}

The largest cluster (\textit{M}) of unvaccinated agents plays a key role by limiting and conditioning the maximum outbreak size.
We study the final outbreak size over the largest cluster, initiated by one infected agent chosen uniformly at random in \textit{M}.
Figure~\ref{fig9} shows that outbreaks for \textit{M}-clusters from opinion dissemination models are larger than those corresponding to
\textit{M}-clusters of same size, but obtained from the percolation approach. Additionally, differences 
can also be observed between both static network and adaptive network opinion dissemination models. 
The origin of such differences should rest on the topological features of the \textit{M}-cluster. 

In order to perform a topological characterization of \textit{M}-clusters, we compute some frequently used quantities such as average degree ($\langle k\rangle$),
 average path length ($\bar{l}$), clustering coefficient ($\bar{C}$) and modularity ($Q$), all as a function of \textit{M}-cluster size.
 Figure~\ref{fig10} compares these topological features for \textit{M}-clusters from opinion dissemination models
 with those corresponding to the equivalent percolation approach, all in the particular case of $\phi = 0.01$ 
 (the same qualitative behavior is obtained for $\phi=0.00$ (Figure Supp2~\ref{figS2}) and $\phi=0.02$ (Figure Supp3~\ref{figS3})). 
 The static network model shows larger $\langle k\rangle$ and $\bar{C}$, while smaller $\bar{l}$ and $Q$, when faced with the equivalent 
 percolation approach results. The same behavior is obtained for the adaptive network model, but in this case all discrepancies with percolation
 approach get smaller. In fact, in this last case, the final networks over which percolation approach is performed, become small-world networks
 as a consequence of the rewiring mechanism performed by the adaptive network model. Then, the percolation threshold of these small-world
 networks results smaller compared with the corresponding to the alternative static regular network.
 Once \textit{M}-cluster size surpasses the percolation threshold, discrepancies between both models and the corresponding percolation approach become smaller.

\textit{M}-clusters obtained by the adaptive model show smaller $\langle k\rangle$ than those for the static model. On the other hand, they also
show smaller $\bar{l}$ and $\bar{C}$, as a consequence of the rewiring mechanism, but an equivalent high modularity $Q$. In summary, \textit{M}-clusters obtained by the
adaptive model are more tree-like than those for the static model (Figure~\ref{fig11}), thereby, this particular feature could lead to fragility 
when faced with a node removal mechanism. 
Following this reasoning, the action of $\phi$ could be interpreted as a mechanism for unvaccinated nodes removal, thus explaining the greater impact of $\phi$ on the adaptive model.

\section*{Summary and Conclusions}

In this work, we argue that people's vaccination behavior are not grounded on a well established scientific knowledge 
allowing them to evaluate infection risks, but on a process of opinion dissemination.
In previous models, the vaccination behavior and the disease spread over the same social network. 
Nowadays, social influence can be exerted without personal contact through current information technologies \cite{Aral2009,Bond2012,Fowler2014}. 
Thus, here we consider two entangled networks comprising personal contacts and opinion non-personal contacts. 
Personal contacts allow infectious disease spreading and also contribute to the process of cultural dissemination. 
Meanwhile, non-personal contacts contribute to disseminate opinions but not to spread an infectious disease.  

We propose two alternative models for dissemination of vaccination behavior between agents placed on a social network.
Both constitute variants of Axelrod's model where vaccination behavior is related to the value attained by a new cultural feature which 
can adopt only two traits: vaccinator or non-vaccinator. Also, this complementary feature is subject to the same dynamics as the other ones.
To the best of our knowledge, this is the first paper to introduce an inhomogeneity in the number of traits $q$ for Axelrod's model.
In addition, our adaptive network model combines two fundamental mechanisms to explain the association of similar individuals: 
social imitation and ``linking by homophily''. In this way, we have obtained clusters of unvaccinated agents 
also satisfying small-world features, widely observed in real social networks \cite{Watts1998,Amaral2000}.

We study the impact of a small anti-vaccination movement over the two variants of our model. 
We show that vaccine refusal spreads more efficiently for the adaptive model in the absence of external intervention.
On the other hand, the impact of public health campaigns, here represented by the external field $\phi$, proved to be very efficient by increasing the average vaccine coverage 
(i.e. reducing unvaccinated population) and, then, reducing the outbreak probability and its final size. 
Moreover, the mitigating effect of $\phi$ has shown to be stronger for the more realistic adaptive model, contrasting with 
its larger efficiency spreading vaccine refusal opinion at $\phi=0$.
These conclusions have been supported through the topological analysis of the largest cluster of unvaccinated (\textit{M}-cluster) on each realization. 
In comparison with the adaptive model, the static model gives place to denser \textit{M}-clusters (higher $\langle k \rangle$) with higher clustering coefficient, 
then more resistant against a mechanism of node removal induced by the external field $\phi$. 
Meanwhile, \textit{M}-clusters from adaptive model show small-world feature, thus reaching larger sizes in absence of external field.
However, they result more fragile against the action of $\phi$ because of their particular topological characteristics.

A strong clustering of unvaccinated agents is apparent when comparing cluster size distributions from our opinion dissemination models with 
those yielded by percolation approach, in which unvaccinated agents are randomly placed over the final networks.
As a direct consequence of this, we observe a marked increase in the outbreak size and its probability of occurrence.
This clustering effect has deeper consequences on the expected herd immunity, even though the recommended vaccine coverage goal ($95\%$ for MMR vaccine) was reached,
which suggests that vaccination goals should also consider the chance of clustering of unvaccinated agents.
Moreover, the clustering effect is magnified under high vaccination coverage regime, in agreement with previous findings \cite{Salathe2008}. 

Our results suggest, on one side, that vaccine availability is not enough to prevent measles outbreaks if it is not 
complemented by consciousness-raising campaigns conducted in order to undermine the harmful action 
of anti-vaccination groups.
On the other side, a better knowledge of the topological features of social networks and their time-dynamic \cite{Eubank2004,Isella2011,Medus2014} would allow improvements in
the modeling of spreading phenomena on them. Finally, the clustering effect should be extensively studied in next experimental tests,
as some authors have begun to do for the case of influenza \cite{Salathe2011,Salathe2014}, in order to provide new insight that effectively lead us to the ultimate eradication
of measles.


\section*{Acknowledgments}
C.O.D. is a member of the ``Carrera del Investigador Cient\'ifico'' CONICET.

%
%
%

\section*{Figure Legends}
%
\begin{figure}[htp]
\caption{
{\bf Social network with personal and non-personal only-opinion contacts.}  Each agent is represented by a node connected with 8 personal
contacts (P) corresponding to its near-neighbors in the 3N+2N square network. The remaining 8 edges are tied to non-personal contacts (O), 
which only participate in the opinion dissemination process, but they are not able to spread an infectious disease. 
}
\label{fig1}
\end{figure}

\begin{figure}[htp]
\caption{
{\bf Unvaccinated agents population as a function of the number of traits $q$ and the external field $\phi$.} Both graphs show the average population
of unvaccinated over 500 runs for each $q$ and $\phi$, under the influence of an anti-vaccination movement representing $1\%$ of total population. 
The results are shown for the static model (A) and the adaptive model (B). The projections over $\phi=0.00$ to $\phi=0.06$ are also shown in both cases. 
}
\label{fig2}
\end{figure}

\begin{figure}[htp]
\caption{
{\bf Sequence of histograms for unvaccinated population under increasing external field strength.} All histograms are normalized. They were obtained performing
500 runs with $q=170$ and $1\%$ of stubborn agents located at the same network sites for each run. The mitigating effect for increasing strength of the external field $\phi$, with
$\phi=0.00$, $0.01$ and $0.02$, is shown in the sequence \textbf{A}, \textbf{B}, \textbf{C} for the static network model and \textbf{D}, \textbf{E}, \textbf{F}
for the adaptive network model.
}
\label{fig3}
\end{figure}

\begin{figure}[htp]
\caption{
{\bf Mitigating effect of $\phi$ on the right tail probability $P(x\geq 25)$ of disease outbreaks involving a population larger than $25$ agents.}  This figure emphasizes the 
greater mitigating effect of $\phi$ on the adaptive model.  $P(x\geq 25)$ was computed through $10000$ simulations of the measles spreading model in each case.
}
\label{fig4}
\end{figure}

\begin{figure}[htp]
\caption{
{\bf Mean outbreak size as function of unvaccinated population.}  Mean outbreak sizes for both opinion dissemination model
($q=170$) were obtained averaging over $10000$ realizations of measles spreading model for each unvaccinated population.
Results from the static model ((\textbf{A}) $\phi=0.00$, (\textbf{C}) $\phi=0.01$, (\textbf{E}) $\phi=0.02$) and from 
the adaptive model ((\textbf{B}) $\phi=0.00$, (\textbf{D}) $\phi=0.01$, (\textbf{F}) $\phi=0.02$) are compared with those 
obtained through the percolation approach.
}
\label{fig5}
\end{figure}

\begin{figure}[htp]
\caption{
{\bf Cluster size distributions of unvaccinated agents.} Cluster size distributions for the static network model ((\textbf{A}) $\phi=0.00$,
(\textbf{C}) $\phi=0.01$, (\textbf{E}) $\phi=0.02$) and the adaptive network model ((\textbf{B}) $\phi=0.00$,
(\textbf{D}) $\phi=0.01$, (\textbf{F}) $\phi=0.02$).  Each distribution comprises the accumulated results of 500 runs of the static or adaptive model with $q=170$ and 
the indicated value of $\phi$. The comparison with percolation is presented and also quantified through the $t$-statistic (see Eq.~\ref{eq4} in the main text) plotted in the inset graphs.
}
\label{fig6}
\end{figure}

\begin{figure}[htp]
\caption{
{\bf Outbreak size distributions for $95\%$ vaccination coverage.}  Static model results
((\textbf{A}) $\phi=0.00$, (\textbf{C}) $\phi=0.01$, (\textbf{E}) $\phi=0.02$) and adaptive model results 
((\textbf{B}) $\phi=0.00$, (\textbf{D}) $\phi=0.01$, (\textbf{F}) $\phi=0.02$) compared with the percolation approach results 
at the same unvaccinated population. Each histogram comprises $10000$ simulations of the measles spreading model.   
}
\label{fig7}
\end{figure}

\begin{figure}[htp]
\caption{
{\bf Cluster size distributions for $95\%$ vaccination coverage.}   Cluster size distributions restricted to about $5\%$ unvaccinated population, 
for the static model ((\textbf{A}) $\phi=0.00$, (\textbf{C}) $\phi=0.01$, (\textbf{E}) $\phi=0.02$) and the adaptive model
((\textbf{B}) $\phi=0.00$, (\textbf{D}) $\phi=0.01$, (\textbf{F}) $\phi=0.02$) compared with the percolation approach results 
at the same unvaccinated population.
}
\label{fig8}
\end{figure}

\begin{figure}[htp]
\caption{
{\bf Comparison of mean outbreak size at the same \textit{M}-cluster size.} Static model ((\textbf{A}) $\phi=0.00$, (\textbf{C}) $\phi=0.01$, (\textbf{E}) $\phi=0.02$) and 
adaptive model ((\textbf{B}) $\phi=0.00$, (\textbf{D}) $\phi=0.01$, (\textbf{F}) $\phi=0.02$) compared with the percolation approach results 
now at the same largest cluster size. Each point corresponds to the average over $10000$ simulations of the measles 
spreading model.
}
\label{fig9}
\end{figure}

\begin{figure}[htp]
\caption{
{\bf Topological description of \textit{M}-clusters for $\phi=0.01$.}  Average degree ($\langle k\rangle$),
 average path length ($\bar{l}$), clustering coefficient ($\bar{C}$) and modularity ($Q$), for both static 
 ((\textbf{A}, (\textbf{C}), (\textbf{E}), (\textbf{G})) and adaptive model (((\textbf{B}, (\textbf{D}), (\textbf{F}), (\textbf{H})))
 compared with those calculated on \textit{M}-clusters of the same size but obtained by the percolation approach.
}
\label{fig10}
\end{figure}

\begin{figure}[ht]
\caption{
{\bf \textit{M}-clusters representation.}  Typical unvaccinated cluster of size $N\sim 200$ for the static network model (\textbf{A}), and 
for the adaptive network model (\textbf{B}), both with external field $\phi=0.01$.
}
\label{fig11}
\end{figure}

%
%
%
%
\section*{Supporting Information Legends}
%
%
\begin{description}
\item{\bf FigS1  \label{figS1} Dependence of unvaccinated fraction with $q$.} 
\item{\bf FigS2. \label{figS2} Topological description of \textit{M}-clusters for $\phi=0.00$.}  Average degree ($\langle k\rangle$),
 average path length ($\bar{l}$), clustering coefficient ($\bar{C}$) and modularity ($Q$), for both static 
 ((\textbf{A}, (\textbf{C}), (\textbf{E}), (\textbf{G})) and adaptive model (((\textbf{B}, (\textbf{D}), (\textbf{F}), (\textbf{H})))
 compared with those calculated on \textit{M}-clusters of the same size but obtained by the percolation approach.
\item{\bf FigS3. \label{figS3} Topological description of \textit{M}-clusters for $\phi=0.02$.}  Average degree ($\langle k\rangle$),
 average path length ($\bar{l}$), clustering coefficient ($\bar{C}$) and modularity ($Q$), for both static 
 ((\textbf{A}, (\textbf{C}), (\textbf{E}), (\textbf{G})) and adaptive model (((\textbf{B}, (\textbf{D}), (\textbf{F}), (\textbf{H})))
 compared with those calculated on \textit{M}-clusters of the same size but obtained by the percolation approach.
\newpage
\item {\bf Epidemiological model description~\label{Supp}}

Measles is a viral infectious disease of the respiratory system. It's a highly contagious airborne disease mainly transmitted by 
close person-to-person contact through Fl\"ugge droplets. Also it is considered one of the leading causes of death among young children \cite{WHOfactsheet},
although it can be contracted at any age by individuals who had not been immunized by vaccination or previous contagion.

In this work, we have modeled the dynamics of measles by means of a simple stochastic compartmental model, in terms of the variables describing the possible 
states that an individual can go through: susceptible ($S$), exposed ($E$), infectious ($I$) and recovered ($R$). 
Only those individuals in infectious state are able to transmit the disease, while recovered individuals acquire permanent immunity. We assume 
a constant population $N$, i.e. without demographic effects, with a social network structure as described in the main text. 
Our susceptible population are children because we assume immunized parents. In addition, we conjecture that measles spreads by
personal contacts only, and these kinds of contacts between children are propitiated mainly by their parents, i.e. 
young children have limited autonomy.
\begin{table}[!h]
\centering
\caption{Transition probabilities for SEIR measles model.}\label{table1}
\begin{tabular}{p{5cm}p{5cm}}
\hline
\hline
State transition & Probability\\
\midrule
$S\rightarrow E$&$1-\left(1-\delta t \:c/\langle k_p\rangle\right)^{z(i)}$\\
\hline
$E\rightarrow I$&$\nu_E \delta t$\\
\hline
$I\rightarrow R$&$\nu_I \delta t$\\
\hline
\hline
\end{tabular}
\end{table}%

The possible transition probabilities between states are defined in Table~\ref{table1}. 
Particular attention deserves the infection probability ($P(S\rightarrow E)$) given that we are considering a contact pattern described by 
the underlying social network, where $z(i)$ corresponds to the number of infectious near-neighbors for the agent $i$. Thus, 
$P(S\rightarrow E)=1-\left(1-\delta t \:c/\langle k_p\rangle\right)^{z(i)}$
takes a different value for each agent $i$, being $c/\langle k\rangle$ the mean contagion probability per edge (we consider $c=2.8$, compatible with known
basic reproductive ratio for measles, and $\langle k_p\rangle=8$ the average degree of personal neighbors for our social network). 

In Table~\ref{table1}, $\nu_E$ and $\nu_I$ are the transition rates from $E$ to $I$ state, and from $I$ to $R$ state, respectively.
Further, transition rates are related with the mean time spent in exposed ($T_E=8$ days, $\nu_E=1/T_E$) and infectious ($T_I=8$ days, $\nu_I=1/T_I$) 
states \cite{Lloyd2001}. 
 
Let $S_t$, $E_t$, $I_t$ and $R_t$ being the populations of susceptible, exposed, infectious and recovered agents at time $t$, respectively. 
Thus, the population of each state is updated at each time step by the following algorithmic prescriptions:
\begin{itemize}
 \item[1. ] Each susceptible agent $i\in S_t$ gets exposed to measles, or not, with the probability given in Table~\ref{table1} by only considering those infectious agents
 $j$ in his neighborhood with $j\in I_t$. The susceptible population is updated for the next time step: $S_{t+\delta t}$.
 \item[2. ] Each exposed agent $i\in E_t$ gets infectious with probability $\nu_E \delta t$. The exposed population is updated for the next time step: $E_{t+\delta t}$.
 \item[3. ] Each infected agent $i\in I_t$ gets recovered with probability $\nu_I \delta t$. The infected population is updated for the next time step: $I_{t+\delta t}$.
 \item[4. ] Update time $t=t+\delta t$ and go to item 1 until $E_t+I_t=0$. 
\end{itemize}
We choose $\delta t=1/2$ day for all our simulations.

\end{description}

\newpage

\begin{figure}[htp]
\captionsetup[subfigure]{labelformat=empty}
        \centering
        \begin{subfigure}[!t]{0.4\textwidth}
                \includegraphics[width=\textwidth,keepaspectratio=true]{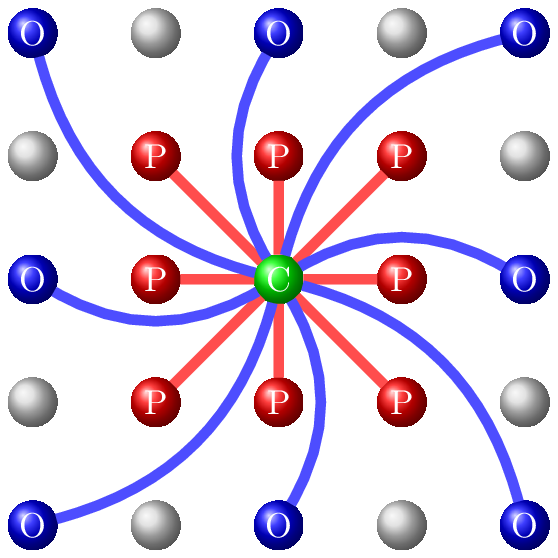}
                \caption{Fig. 1}
        \end{subfigure}%
        \vspace{0.5cm}
        
        \begin{subfigure}[b]{1.0\textwidth}
        \centering
                \includegraphics[width=\textwidth,keepaspectratio=true]{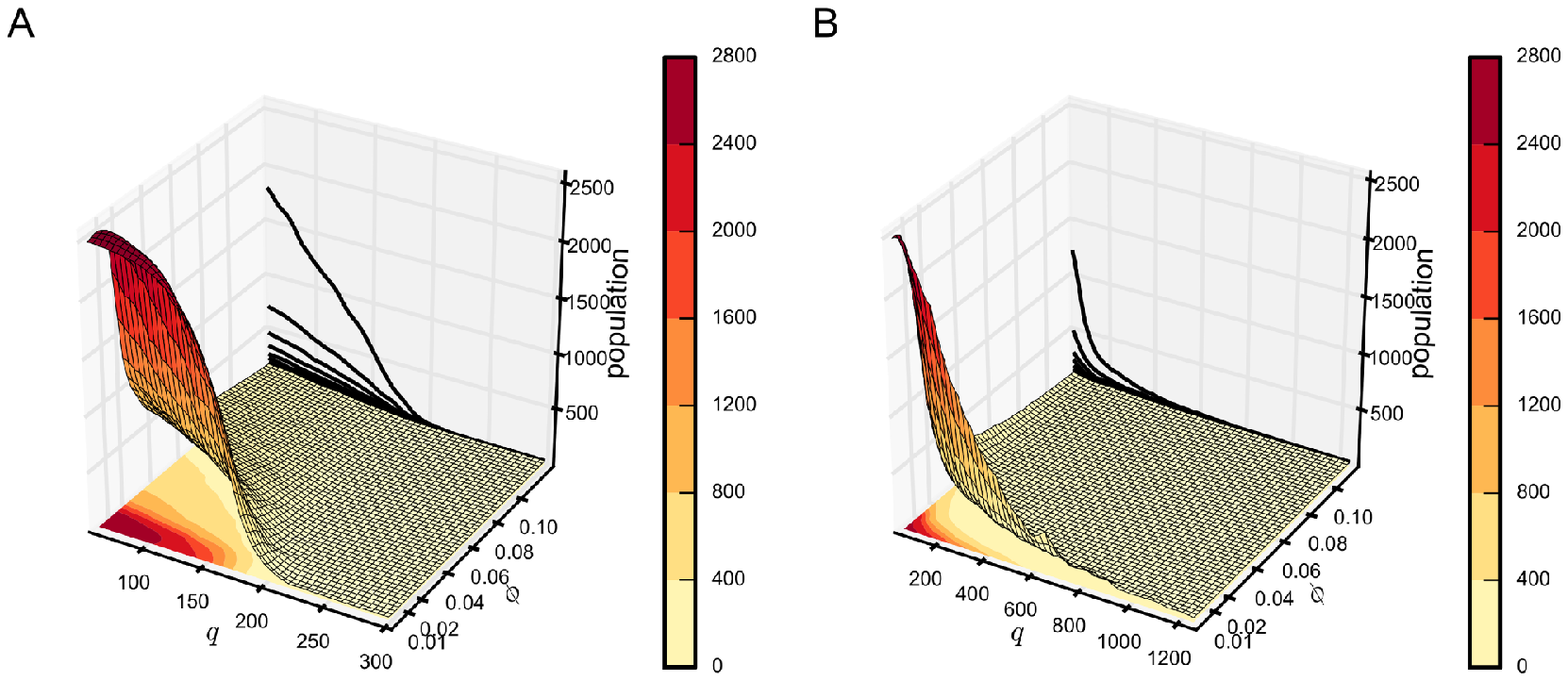}
                \caption{Fig. 2}
        \end{subfigure}
\end{figure}

\begin{figure}[!t]
\captionsetup[subfigure]{labelformat=empty}
        \centering
        \begin{subfigure}[t]{0.6\textwidth}
                \includegraphics[width=\textwidth,keepaspectratio=true]{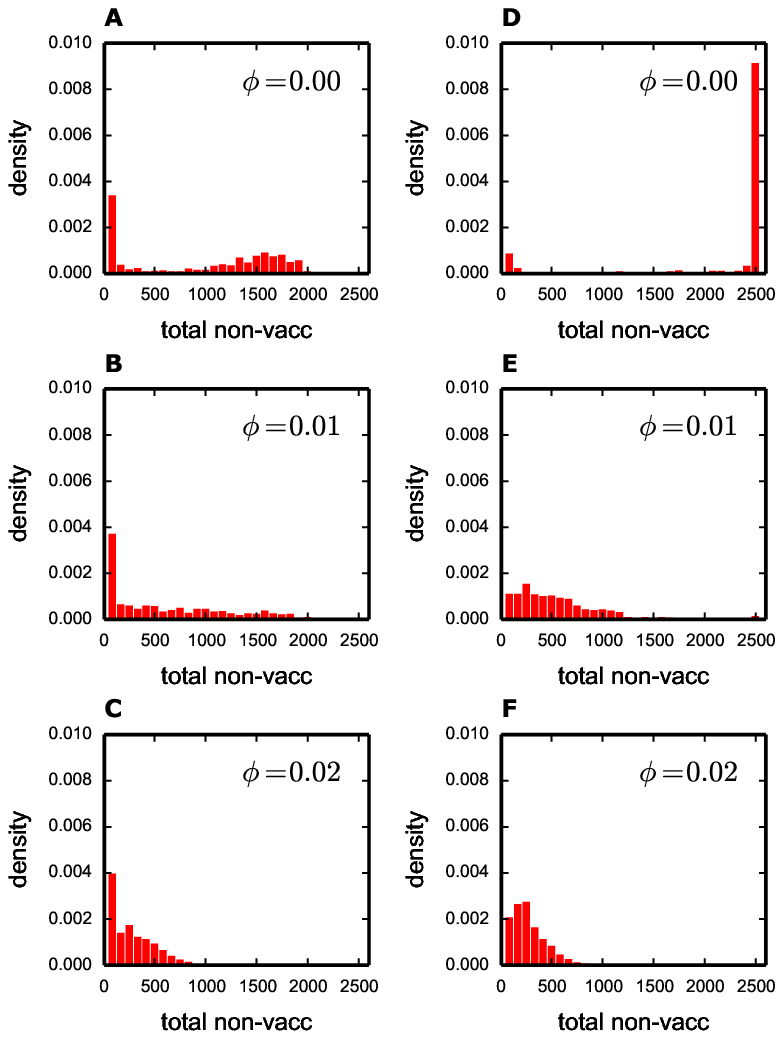}
                \caption{Fig. 3}
        \end{subfigure}%
        
        \begin{subfigure}[t]{0.5\textwidth}
                \includegraphics[width=\textwidth,keepaspectratio=true]{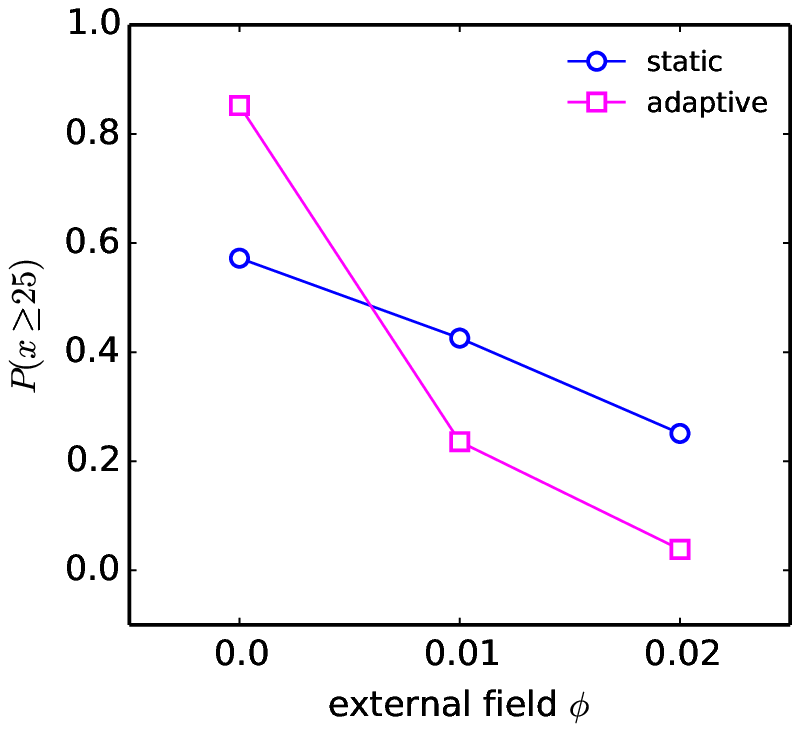}
                \caption{Fig. 4}
        \end{subfigure}
\end{figure}

\begin{figure}[!t]
\captionsetup[subfigure]{labelformat=empty}
        \centering
        \begin{subfigure}[t]{0.8\textwidth}
                \includegraphics[width=\textwidth,keepaspectratio=true]{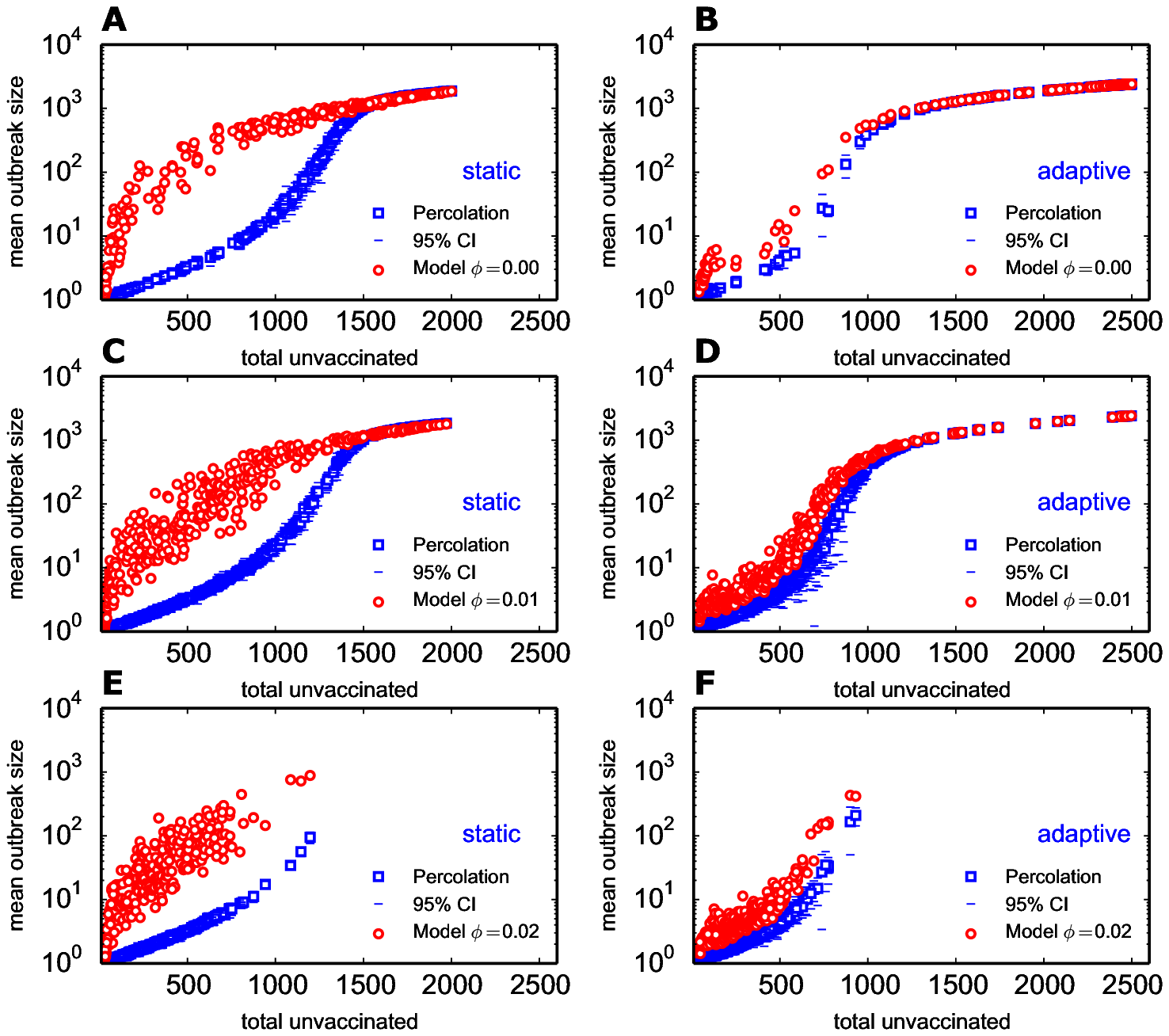}
                \caption{Fig. 5}
        \end{subfigure}%
        
        \begin{subfigure}[b]{0.7\textwidth}
                \includegraphics[width=\textwidth,keepaspectratio=true]{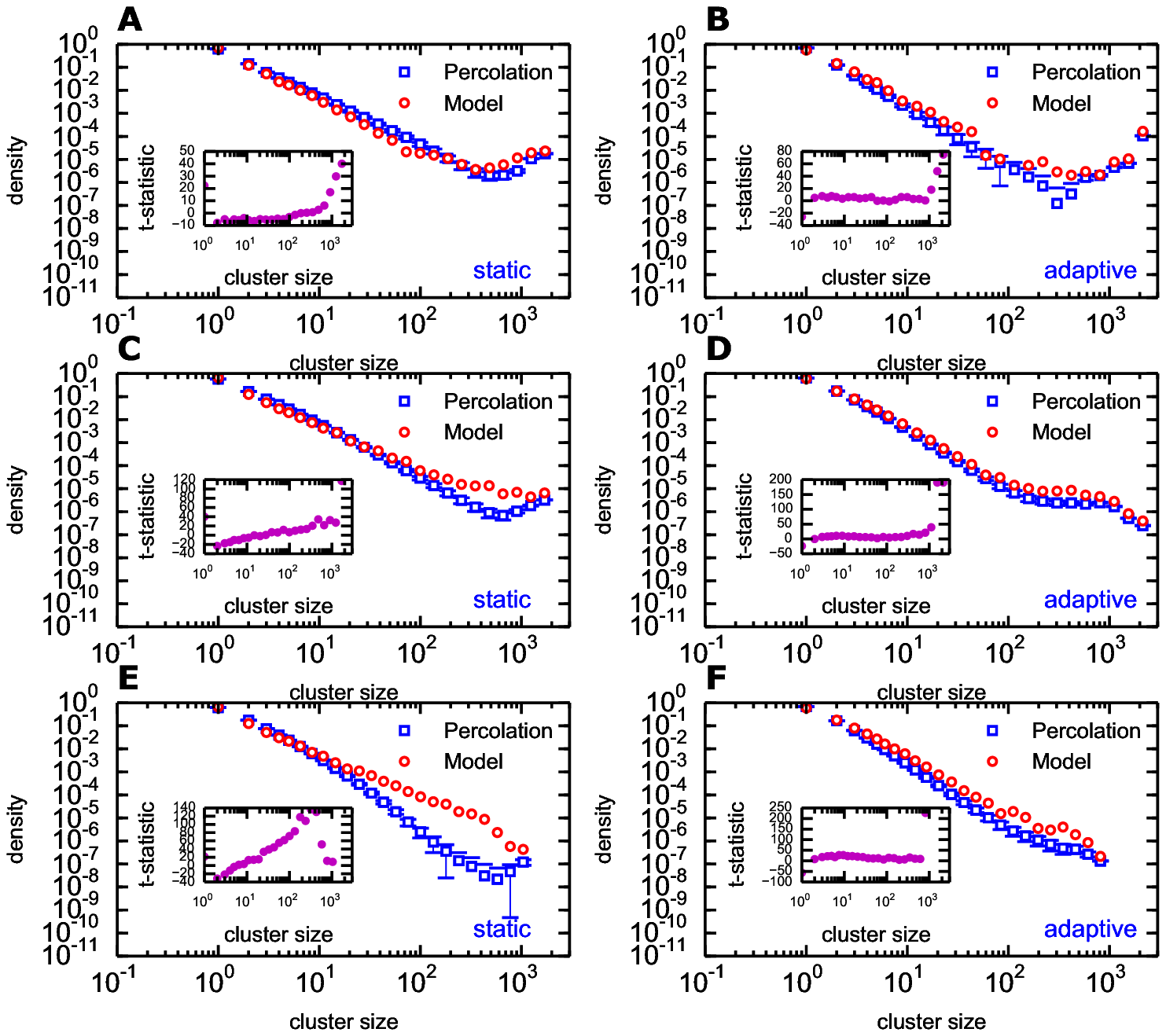}
                \caption{Fig. 6}
        \end{subfigure}
\end{figure}

\begin{figure}[!t]
\captionsetup[subfigure]{labelformat=empty}
        \centering
        \begin{subfigure}[t]{0.8\textwidth}
                \includegraphics[width=\textwidth,keepaspectratio=true]{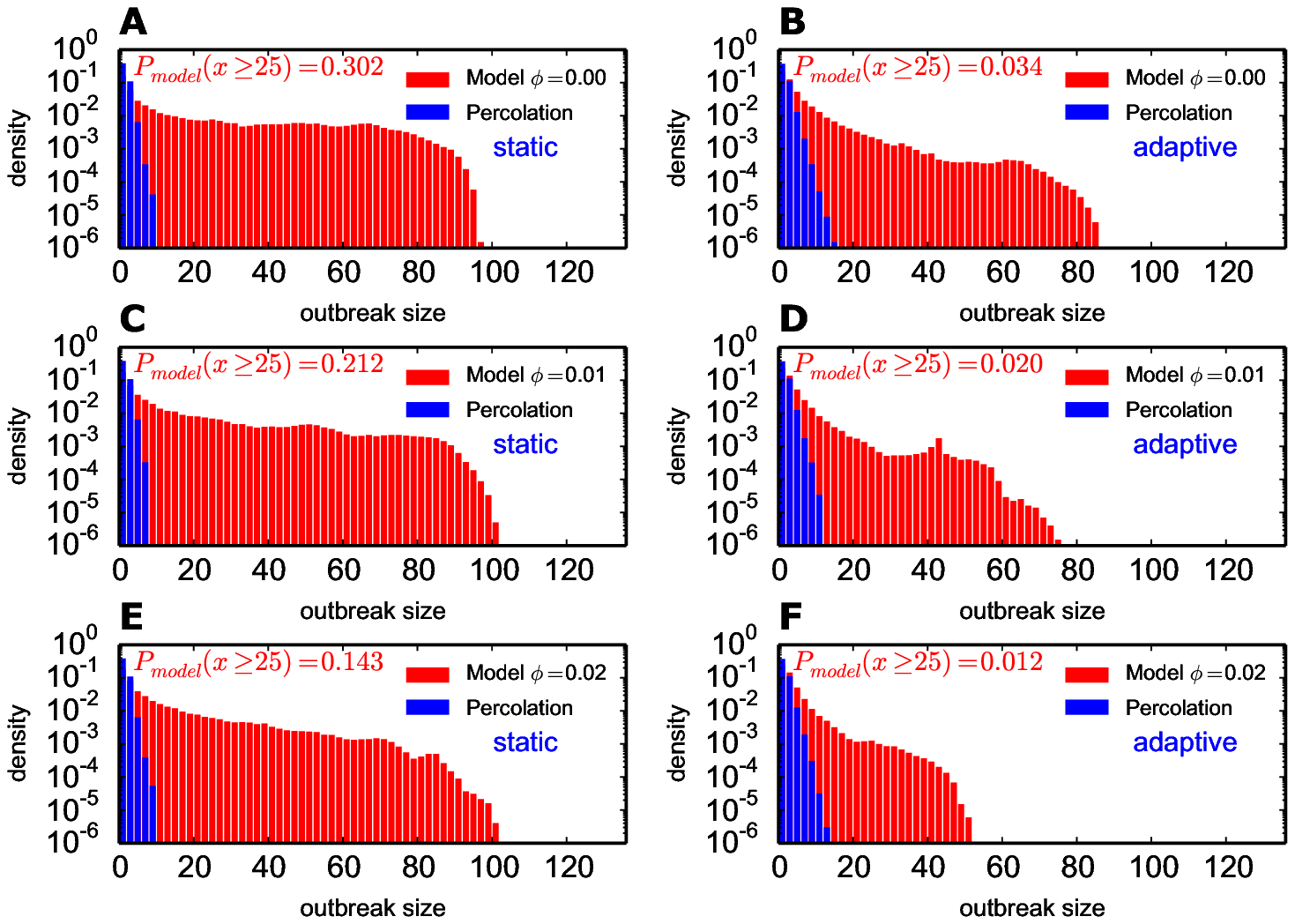}
                \caption{Fig. 7}
        \end{subfigure}%
        
        \begin{subfigure}[b]{0.8\textwidth}
                \includegraphics[width=\textwidth,keepaspectratio=true]{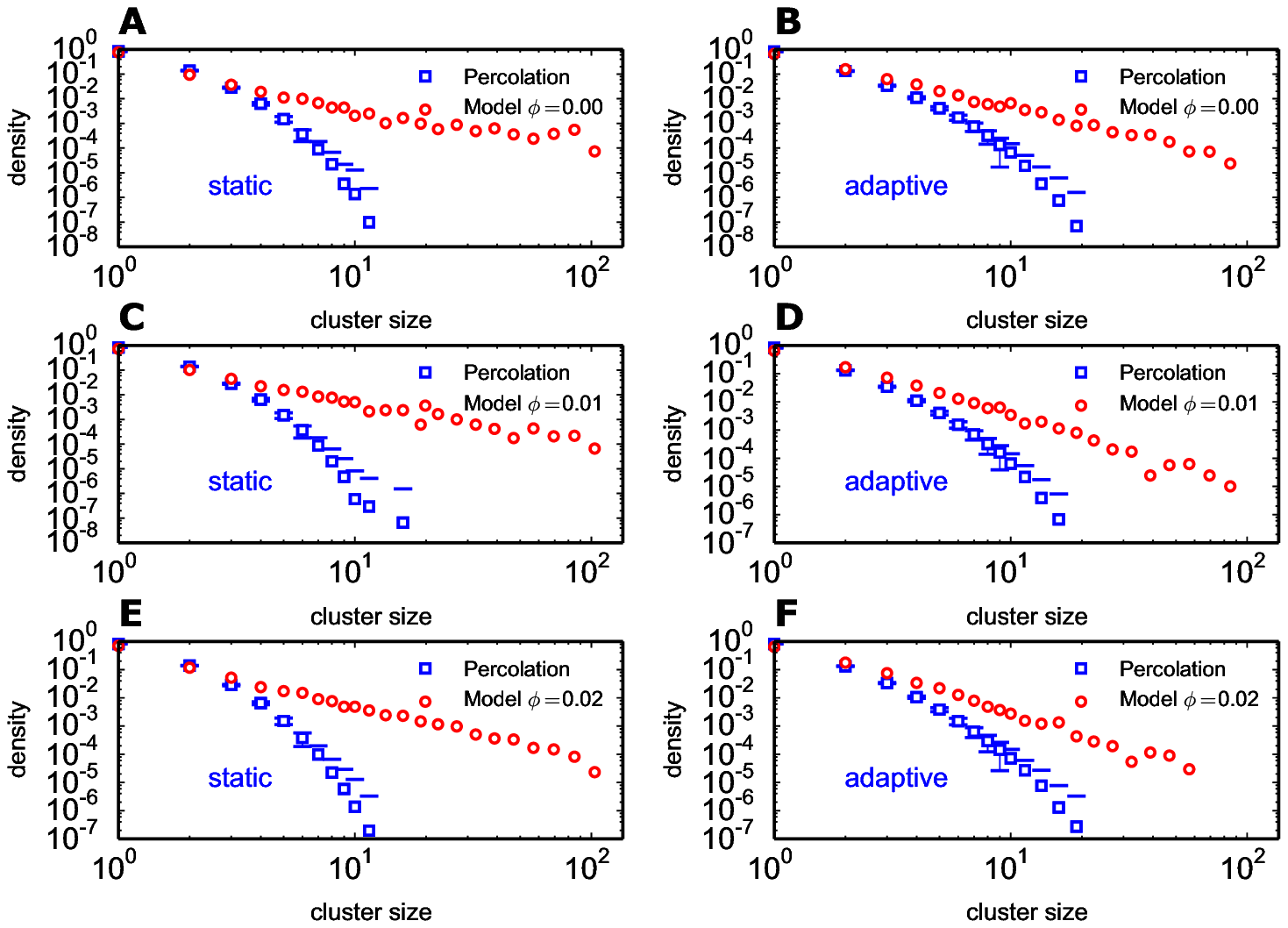}
                \caption{Fig. 8}
        \end{subfigure}
\end{figure}

\begin{figure}[!t]
\captionsetup[subfigure]{labelformat=empty}
        \centering
        \begin{subfigure}[t]{0.7\textwidth}
                \includegraphics[width=\textwidth,keepaspectratio=true]{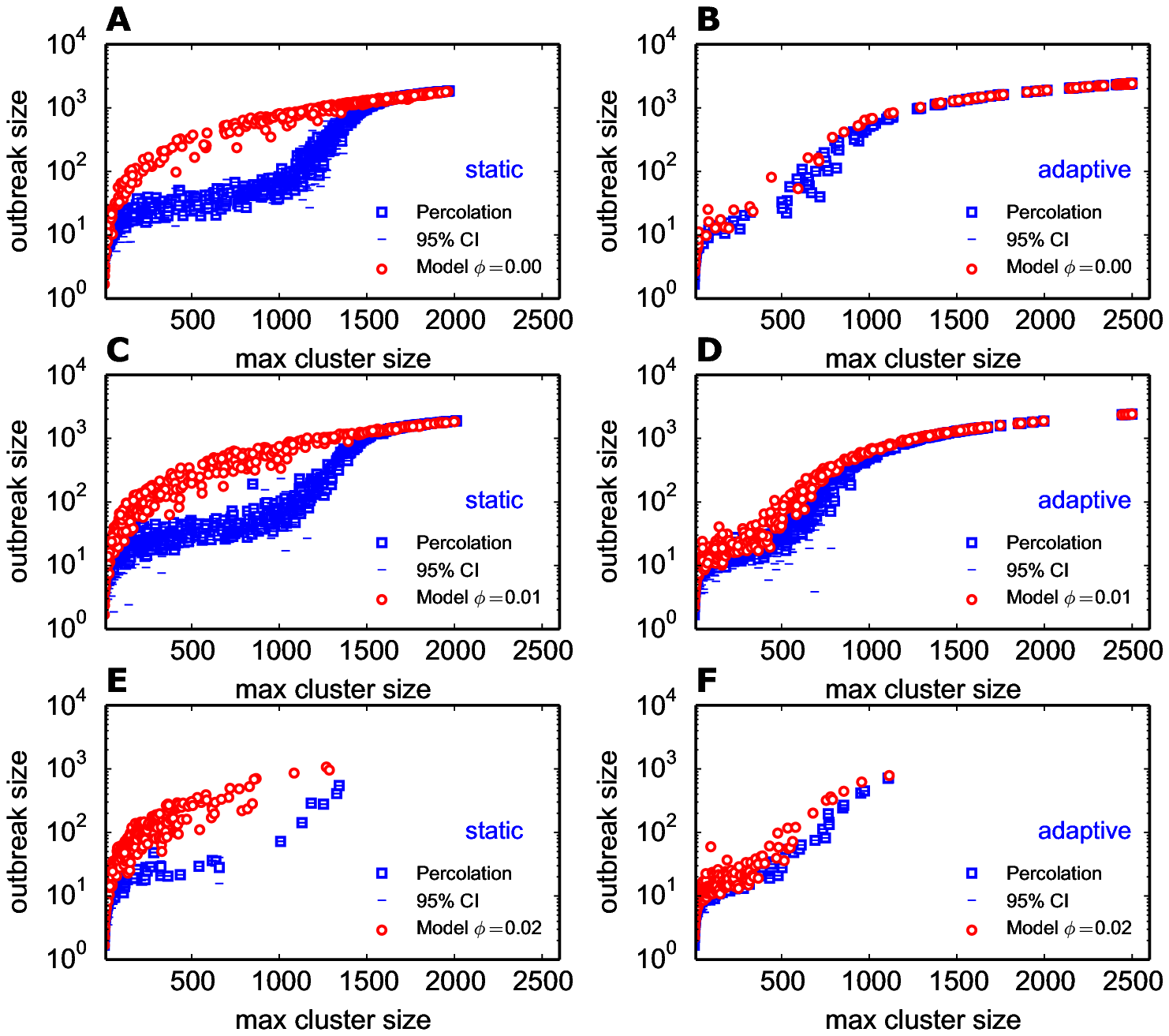}
                \caption{Fig. 9}
        \end{subfigure}%
        
        \begin{subfigure}[b]{0.7\textwidth}
                \includegraphics[width=\textwidth,keepaspectratio=true]{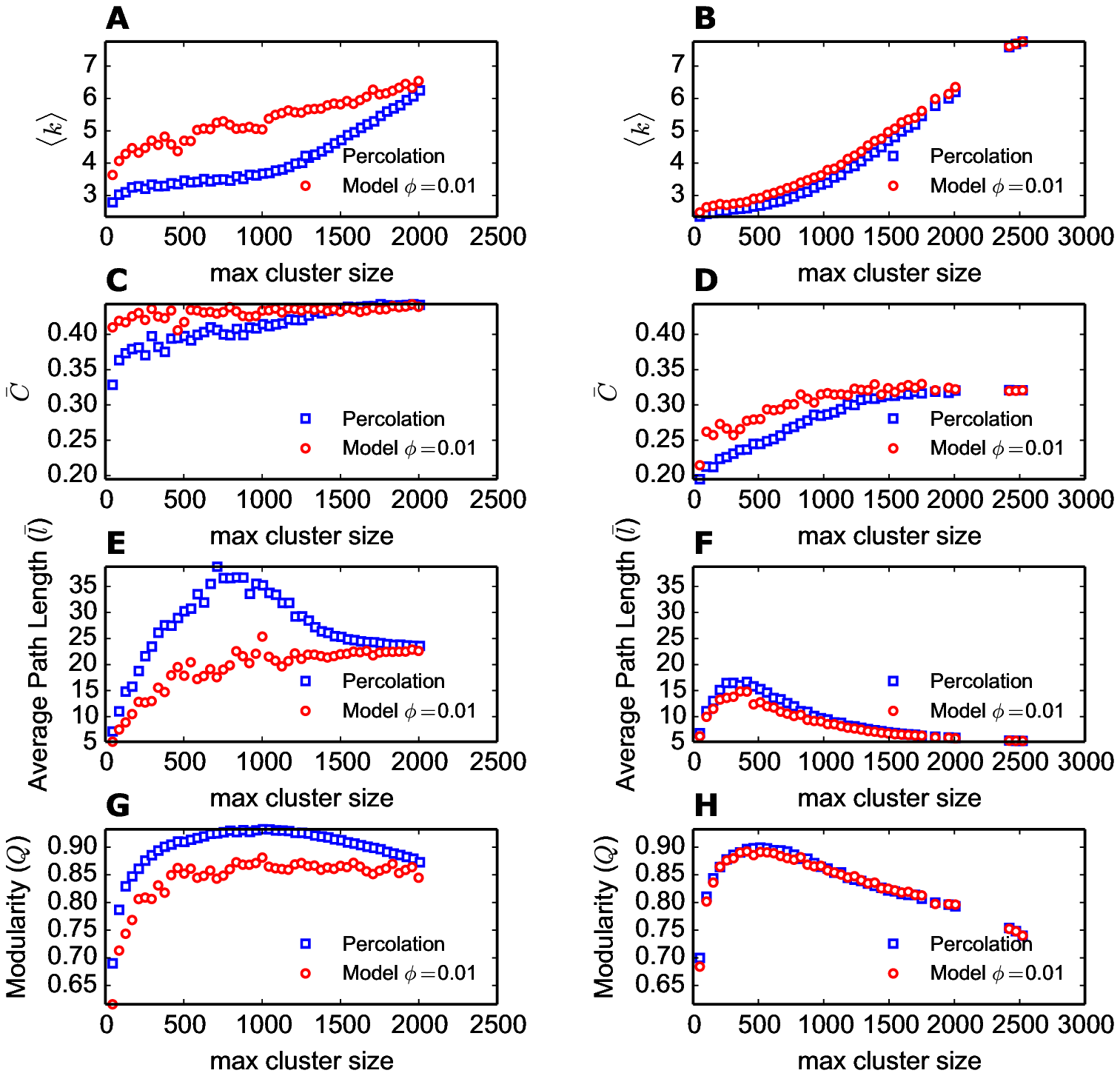}
                \caption{Fig. 10}
        \end{subfigure}
\end{figure}

\begin{figure}[!t]
\captionsetup[subfigure]{labelformat=empty}
        \centering
        \begin{subfigure}[t]{0.8\textwidth}
                \includegraphics[width=\textwidth,keepaspectratio=true]{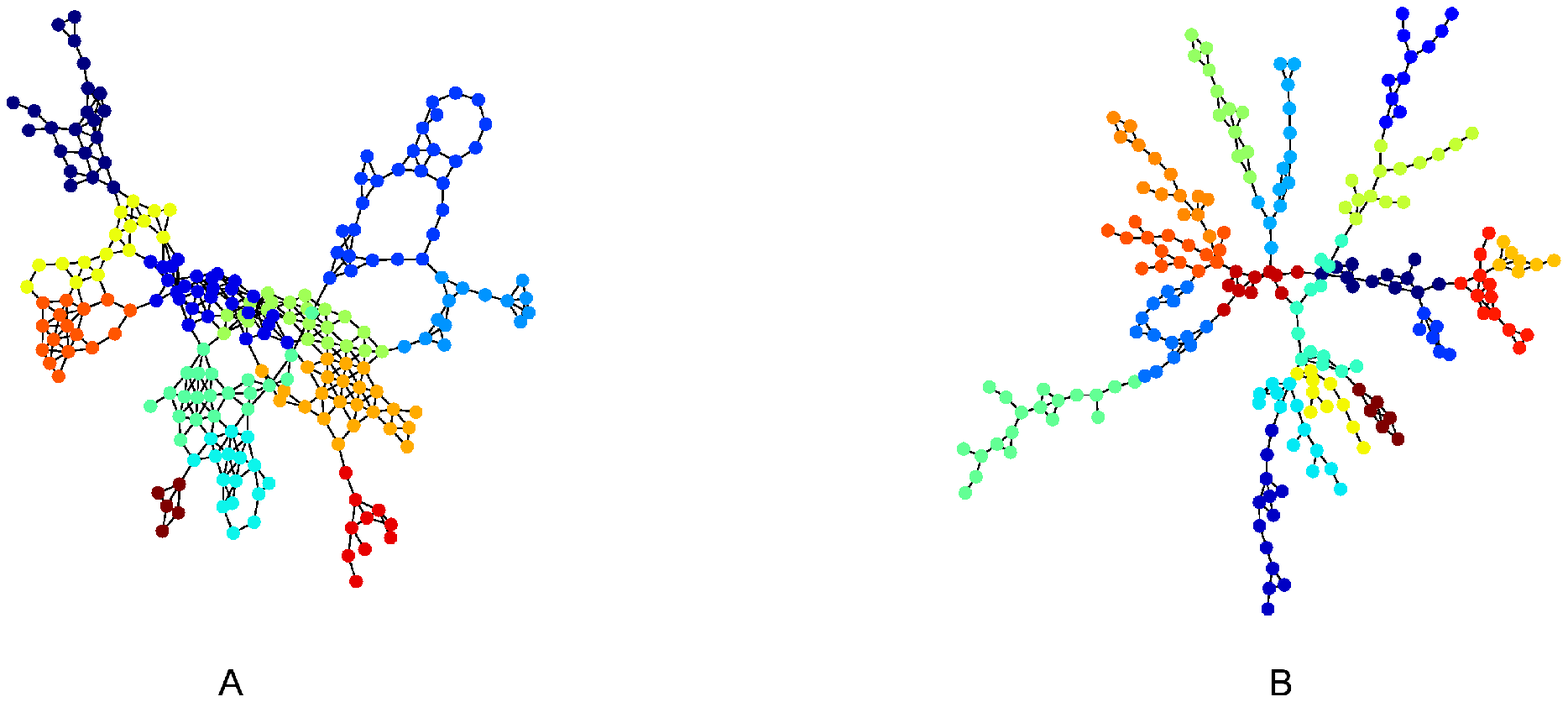}
                \caption{Fig. 11}
        \end{subfigure}%
        
        \begin{subfigure}[b]{0.8\textwidth}
                \includegraphics[width=\textwidth,keepaspectratio=true]{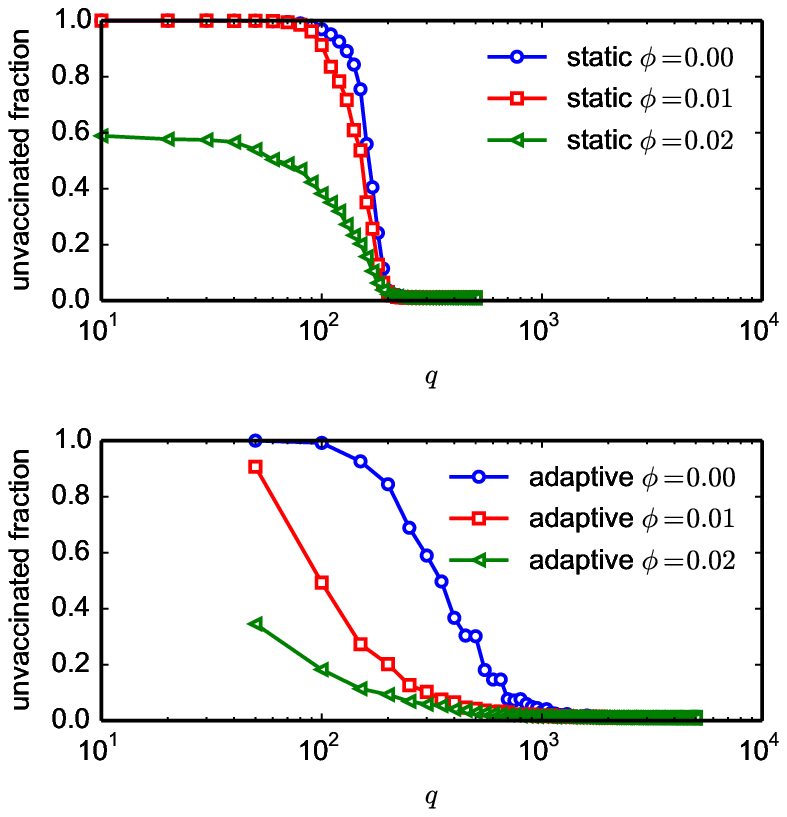}
                \caption{Fig. S1}
        \end{subfigure}
\end{figure}

\begin{figure}[htp]
\captionsetup[subfigure]{labelformat=empty}
        \centering
        \begin{subfigure}[t]{0.7\textwidth}
                \includegraphics[width=\textwidth,keepaspectratio=true]{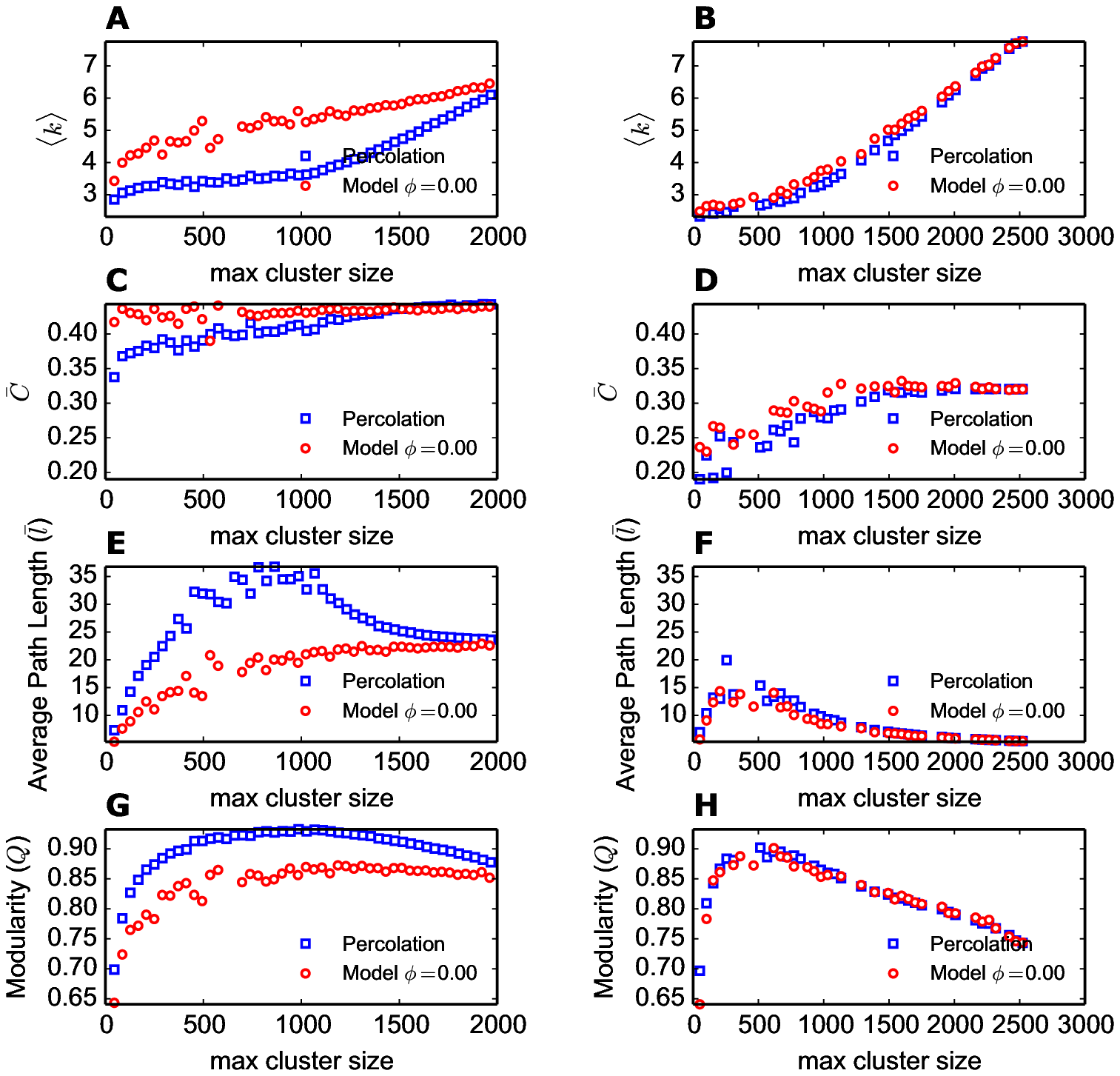}
                \caption{Fig. S2}
        \end{subfigure}%
        
        \begin{subfigure}[b]{0.7\textwidth}
                \includegraphics[width=\textwidth,keepaspectratio=true]{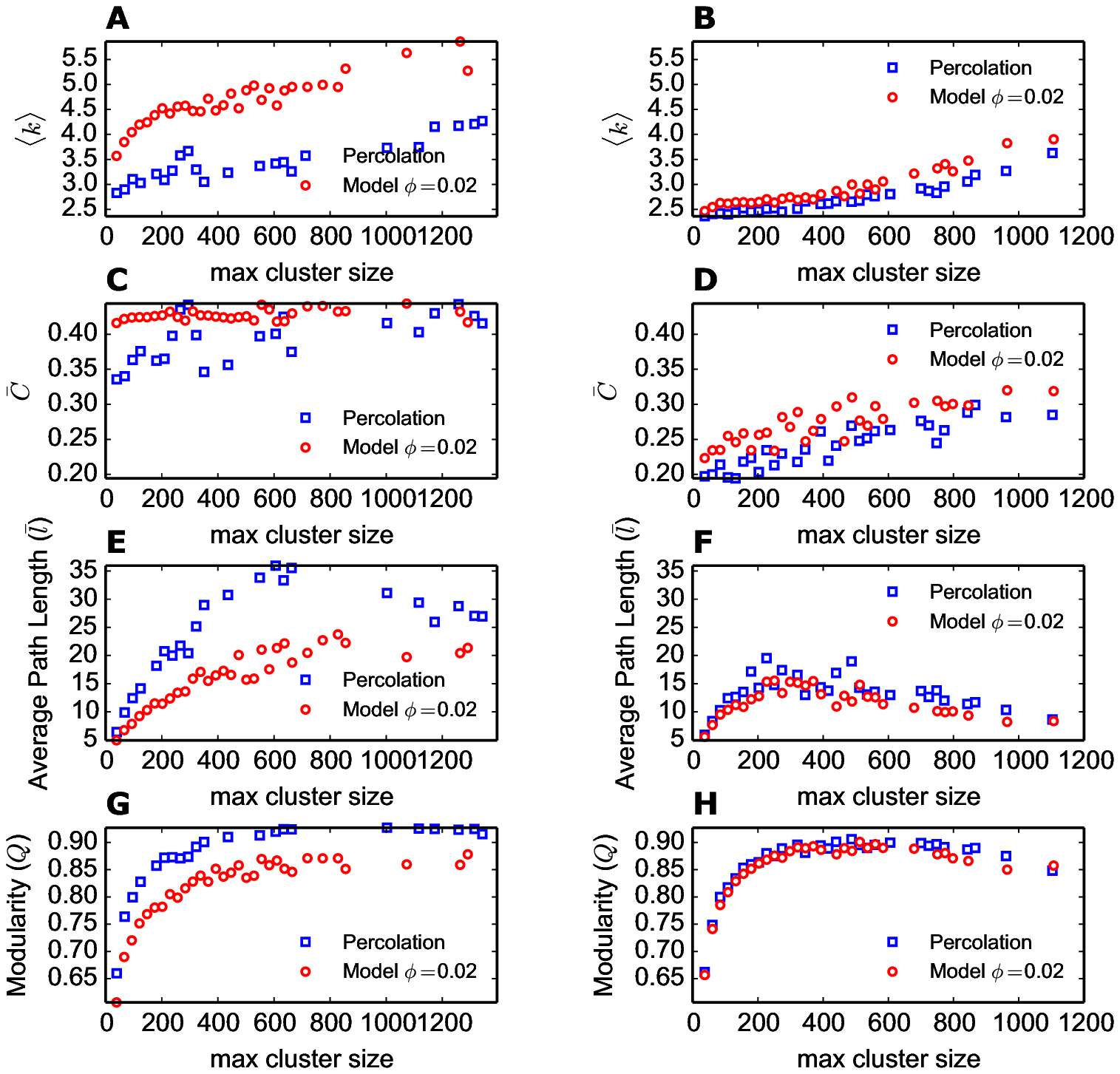}
                \caption{Fig. S3}
        \end{subfigure}
\end{figure}

\end{document}